# Understanding the Optoelectronic Processes in Colloidal 2D Multi-Layered MAPbBr$_3$ Perovskite Nanosheets: Funneling, Recombination and Self-Trapped Excitons

*André Niebur, Eugen Klein, Rostyslav Lesyuk, Christian Klinke, and Jannika Lauth\**

A. Niebur, Prof. Dr. J. Lauth

Institute of Physical Chemistry and Electrochemistry, Leibniz University Hannover, Callinstr. 3a, D-30167 Hannover, Germany

E-mail: jannika.lauth@uni-tuebingen.de

A. Niebur, Prof. Dr. J. Lauth

Cluster of Excellence PhoenixD (Photonics, Optics, and Engineering – Innovation Across Disciplines), Hannover, Germany

Dr. E. Klein, Dr. R. Lesyuk, Prof. Dr. C. Klinke

Institute of Physics, University of Rostock, Albert-Einstein-Straße 23, 18059 Rostock, Germany

Dr. R. Lesyuk

Pidstryhach Institute for Applied Problems of Mechanics and Mathematics of NAS of Ukraine, Naukowa Str. 3b, 79060 Lviv, Ukraine

Prof. Dr. C. Klinke

Department Life, Light & Matter, University of Rostock, Albert-Einstein-Strasse 25, 18059 Rostock, Germany

Prof. Dr. C. Klinke

Department of Chemistry, Swansea University – Singleton Park, Swansea SA2 8PP, United Kingdom

Prof. Dr. J. Lauth

Laboratory of Nano and Quantum Engineering (LNQE), Leibniz University Hannover, Schneiderberg 39, D-30167 Hannover, Germany

Prof. Dr. J. Lauth

Institute of Physical and Theoretical Chemistry, University of Tübingen, Auf der Morgenstelle 18, D-72076 Tübingen, Germany

Quasi two-dimensional (2D) colloidal synthesis made quantum confinement readily accessible in perovskites, generating momentum in perovskite LED research and lasing. Ultrathin perovskites exhibit charge transport properties beneficial for solar cells and the combination of layers with different thickness directs charge carriers toward thicker layers with a smaller bandgap. However, detailed knowledge about the mechanisms by which excitons and charge carriers funnel and recombine in these structures is lacking. Here, we characterize colloidal 2D methylammonium lead bromide (MAPbBr$_3$) Ruddlesden-Popper perovskites with a broad combination of layers ($n$ = 3 to 10, and bulk fractions with $n > 10$) by femtosecond transient absorption spectroscopy and time-resolved photoluminescence. We find that second- and third-order processes dominate in MAPbBr$_3$ nanosheets, indicating exciton-exciton annihilation (EEA) and Auger recombination. Long-lived excitons in thin layers (e.g., $n = 5$, $E_b = 136$ meV) funnel into high *n* within 10–50 ps, which decreases their exciton binding energy below $k_B T$ and leads to radiative recombination. Parallel and consecutive funneling compete with trapping processes, making funneling an excellent tool to overcome exciton self-trapping when high-quality *n-n* interfaces are present. Free charge carriers in high-*n* regions on the other hand facilitate radiative recombination and EEA is bypassed, which is desirable for LED and lasing applications.

## 1. Introduction

Solar photons reaching earth could cover the global energy consumption more than 8000-fold.[1] We therefore need to make this abundant energy source more accessible through affordable solar cells. Such devices based on cost efficient perovskites exhibit steadily increasing efficiencies exceeding 26 % (or 33 % in tandem solar cells with Si).[2] Besides, perovskites are appealing for many optoelectronic applications, since they show a tunable bandgap[3,4], near unity photoluminescence quantum yield (PLQY)[5] and sharp emission[6]. In this context, the underlying photophysics in colloidal 2D MAPbBr$_3$ nanosheets, especially on an ultrafast timescale, are crucial for energy conversion.

Perovskites have the formula ABX$_3$, where A is a small cation, most prominently methylammonium (MA, CH$_3$NH$_3^+$) or Cs$^+$, B is a metal, and X is a halogen, forming a corner-connected [BX$_6$]$^{4-}$ octahedron. The small cations are located in the gaps of the B–X structure. When introducing a large organic cation A′, perovskites can form layered structures with a A′$_2$A$_{n-1}$B$_n$X$_{3n+1}$ stoichiometry known as quasi-2D Ruddlesden–Popper perovskites (RPP) with $n$ being the number of octahedron layers, becoming less rigid for smaller $n$.[7] The thinnest RPP ($n$ = 1) is a single layer of B–X octahedrons, which is stabilized by the large organic cations on both sides. Research focused on controlling the number of B–X octahedron layers $n$ is receiving increasing interest, since controlling $n$ allows additional tailoring of optoelectronic properties into the direction of the design of complex band alignments.[8] Among others, studies on the (octahedral) layer numbers have been published for perovskites of the types MAPbI$_3$,[4,9–11] MAPbBr$_3$,[4,12,13], MAPbCl$_3$,[4] CsPbBr$_3$[14,15] and lead free alternatives including MASnI$_3$, MAGeI$_3$, and CsSnI$_3$.[16,17] This series of materials covers the full visible spectrum from the UV (MAPbCl$_3$) to the NIR (MAPbI$_3$) and the structures have proven to be efficient absorbers and emitters for solar cells and in LEDs.[18,19] Tin-based perovskites offer a highly-efficient lead-free material platform for applications, however they also suffer from a rather high degradation.[20,21]

By reducing the layers of 2D perovskites to a single-digit number of crystal units, an atomically precise confinement is introduced, which leads to optoelectronic properties being mainly dependent on the material thickness with lateral sizes often playing a minor role. Furthermore, the properties of these ultrathin perovskites reflect the interplay between quantum confinement and dielectric confinement. The latter reduces Coulomb screening of excitons in nanosheets with a reduced thickness due to a dielectric mismatch with the surrounding medium and effectively increases the exciton binding energy.[22,23] Blancon *et al.* have shown for MAPbI$_3$ RPPs that with reduced thickness, the bandgap, exciton binding energy and the effective mass of the excitons increases. They attribute this behavior to band mixing and non-parabolic effects.[24] Milot *et al.* reported a 25-fold lower carrier mobility with decreasing $n$ in MAPbI$_3$ RPPs, which they associated with a higher exciton binding energy caused by stronger charge carrier confinement.[25] In the context of solar cells, an appealing strategy is to combine different perovskite layer thicknesses to exploit smaller exciton binding energies in high-$n$ RPPs with the directional carrier motion due to band alignment. The detailed charge carrier dynamics of mixed- and multi-$n$ RPPs provide information on ways to refine these structures. For example it has been demonstrated, that excitons in $CsPbBr_3$ RPPs can be trapped by uncoordinated $Pb^{2+}$ on the way to high-$n$ regions, which is suppressed by methanesulfonate treatment.[15] Besides, it is crucial whether the crystal contact between different layers is epitaxial. According to Li *et al.*, the *n-n* interface determines how quickly charge carriers are funneled into high-$n$ regions, which was shown for a $MAPbI_3$ RPP with $n = 3, 4, 5$, and bulk.[9] The authors found that electrons funnel with $\tau < 500$ fs into the bulk if the *n-n* contact is epitaxial, whereas $\tau > 290$ ps is described for a highly defective interface. Recently, Oddo *et al.* showed evidence for funneling of energy rather than charge carriers in mixed-$n$ $CsPbBr_3$ RPPs, which they supported with femtosecond transient absorption spectroscopy (TAS).[26]

Here, we focus on studying Auger and funneling processes in colloidal 2D multi-layered $MAPbBr_3$ RPPs, since non-radiative Auger recombination limits the efficiency in nanocrystal LEDs and shortens the lifetime of gain, therefore preventing efficient stimulated emission and lasing.[27–29] We apply steady-state absorption, TAS, PL, time-resolved PL (time-correlated single photon counting, TCSPC) and support our findings with material characterizing methods including TEM and XRD. The 2D multi-$n$ perovskites investigated here are synthesized based on a previously published method by Klein *et al.* and exhibit a layer thickness of $n = 3, 5, 7, 8, 10$, and bulk ($n > 10$) in different thickness combinations.[4] We apply Elliott theory for steady-state absorption[30] and an extended time-resolved Elliott model for the exited state absorption[31] to determine the transient number of excited charge carriers $N_e$. We find that $N_e$ decays with mixed second- and third-order kinetics, which is indicative of bi-molecular radiative recombination, EEA and a three-body Auger process. Most importantly we show that funneling is neither strictly parallel nor consecutive and carriers can accumulate in bulk-like or intermediate regions, from where they recombine radiatively. With increased photoexcitation fluences, we find evidence for self-trapping of excitons which in turn start to compete with the funneling. Our results not only illustrate which different ultrafast processes can occur in colloidal multi-$n$ $MAPbBr_3$ RPPs, but also help to understand and describe how funneling can be used for e.g. directed energy transfer, which is important for highly efficient solar cells and LEDs.

## 2. Results and Discussion

### 2.1. Structural Characterization of Multi-Layered MAPbBr$_3$

2D MAPbBr$_3$ perovskites were synthesized by first preparing PbBr$_2$ nanosheets, which were converted into MAPbBr$_3$ by a reaction with methylammonium bromide.[4] The conversion was performed at 160 °C in diphenyl ether and a mixture of dodecylamine and trioctylphosphine as ligands. Targeted layer numbers (3, 5, 7, 8, 10, and bulk) were obtained by using oleylamine instead of dodecylamine, and by varying the reaction temperature, time, and concentration of precursors[4] (see SI for detailed syntheses). **Figure 1**a–d shows TEM images of predominantly rectangular MAPbBr$_3$ perovskite nanosheets with lateral sizes between 50 nm and 5 μm. The sample containing $n$ = 5, 7, 8, 10, and bulk (Fig. 1d) is shown in twice the magnification with respect to the others due to its smaller size (a TEM cutout of the same size as for the other samples is shown in Figure S1 in the SI). The layer numbers were validated by UV/vis and PL spectroscopy, showing distinct features for the respective $n$ and XRD confirming the cubic structure (Pm$\bar{3}$m) and "superstacks". Figure 1e shows multi-layered MAPbBr$_3$ nanosheet UV/Vis and PL spectra. The mixtures, e.g. the sample containing 5, 7 layers and a bulk fraction, has one distinct absorbance maximum and two shoulders at 2.76 eV, 2.64 eV and 2.38 eV, resp., with corresponding PL features at 2.72 eV, 2.62 eV and 2.36 eV (main peak). Accordingly, the sample which exclusively contains bulk-like MAPbBr$_3$ nanosheets (black), shows a single absorbance maximum at 2.38 eV and one PL maximum at 2.35 eV. Spectral features and respective layer numbers of all samples are summarized in **Table 1**.

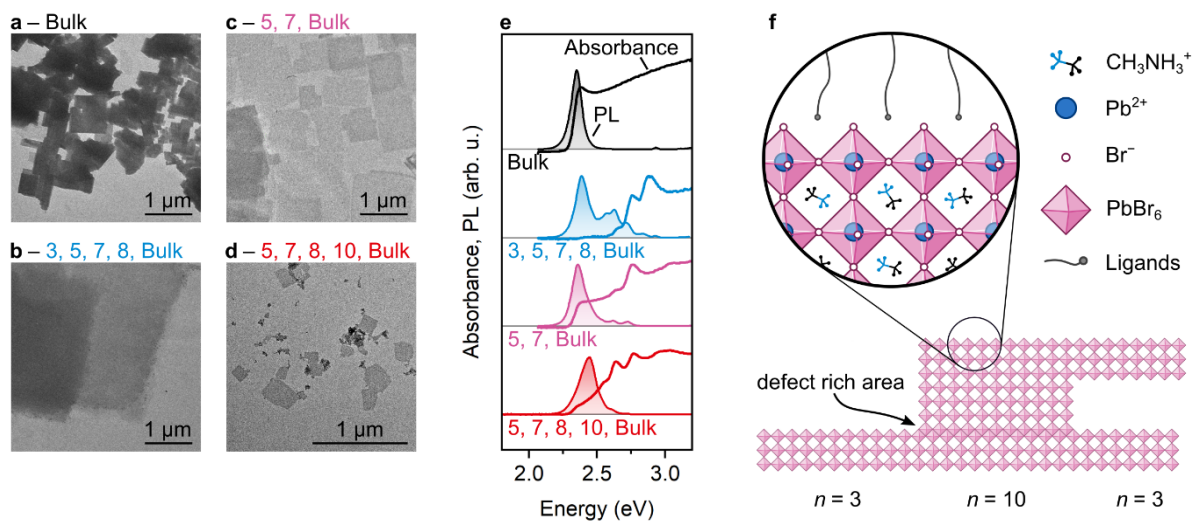

**Figure 1.** Colloidal 2D MAPbBr$_3$ perovskites exhibit a distinct thickness with specific spectral features. (**a** – **d**) TEM images of four samples with bulk-like and/or multi-layered nanosheets, exhibiting either rectangular or irregular lateral shapes, with the sample in (**d**) being considerably smaller than the other samples and showing the characteristic optical

properties. (**e**) Nanosheets absorb and emit at specific spectral positions, which we assign to precise layer numbers, summarized in Table 1. (**f**) Individual nanosheets contain different layer numbers at the same time, allowing for the transfer of energy and electrons.

**Table 1.** Maxima and shoulders in absorption and PL spectra as well as PLQY of colloidal 2D MAPbBr$_3$ perovskites with different composition of layer numbers.

| Sample ($n$ =) | Spectral features (eV [nm]) | | PLQY |
|---|---|---|---|
| Bulk | Absorbance | 2.38 [521] | 0.2 % |
| | PL | 2.35 [528] | |
| 3, 5, 7, 8, Bulk | Absorbance | 2.88, 2.75, 2.64, —, 2.42 [430, 451, 470, —, 512] | 3.6 % |
| | PL | 2.84, 2.72, 2.63, 2.57, 2.38 [437, 456, 471. 482, 521] | |
| 5, 7, Bulk | Absorbance | 2.76, 2.64, 2.38 [449, 470, 521] | 14.3 % |
| | PL | 2.72, 2.62, 2.36 [456, 473, 525] | |
| 5, 7, 8, 10, Bulk | Absorbance | 2.77, 2.64, 2.55, 2.46$^{a)}$, 2.33 [448, 470, 486, 504, 532] | 31 % |
| | PL | 2.73, 2.61, 2.55$^{b)}$, 2.45, 2.31$^{b)}$ [454, 475. 486, 506, 537] | |

$^{a)}$ Corresponding ground state bleach in TAS measurements; $^{b)}$ Determined with the 2$^{nd}$ derivative of the PL spectrum.

The assignment of *n* for the samples discussed here is supported by powder XRD data (see **Figure 2**). The bulk pattern is based on crystallographic data of Elbaz et al.[32] and simulated patterns in grey are obtained according to the method described by Klein et al.[4] Colloidal layered perovskite nano- and micro-crystals show periodic, nearly equidistant XRD intensity patterns plotted *vs.* 2θ in the diffractograms. The position of the reflexes can provide information on the magnitude of the preferential orientation (texture effect) and stacking periodicity. Additionally, the relative intensity of the reflexes yields information on the number of monolayers and thus the organic spacer thickness. We find the best match to the experimental data is provided by *n* values of 3, 5, and 5, for the three multi-layered samples, resp. and an organic spacer thickness of nearly 18 Å, which correlates well with the length of the dodecylamine spacer (≈ 16.5 Å) and implying an intercalated ligand double layer. The *n* found by XRD analysis correspond to the lowest *n* values in the respective sample evidenced by steady-state absorption and TAS. The higher-*n* pattern from the mixed samples could not be clearly recognized due to their low intensity. In the "red" sample (5-7-8-10-bulk), very sharp reflexes at 17° and 34° 2θ were detected (Figure 2). These reflexes represent evidence for thick structures in the sample (up to 90 nm according to the Scherrer equation) and are attributed to the stacking of the layered nanosheets into thicker superstacks. An example of such a superstack is shown in Figure S2, confirming a superstack height of 70 nm to 120 nm. In the bulk sample, reflexes at 14.9°, 21.2°, 30.1°, 33.8°, 37.1°, 43.2°, 45.9° and 62.7° 2θ correspond well with the crystallographic planes (100), (110), (200), (210), (211), (220), (300), and (400).

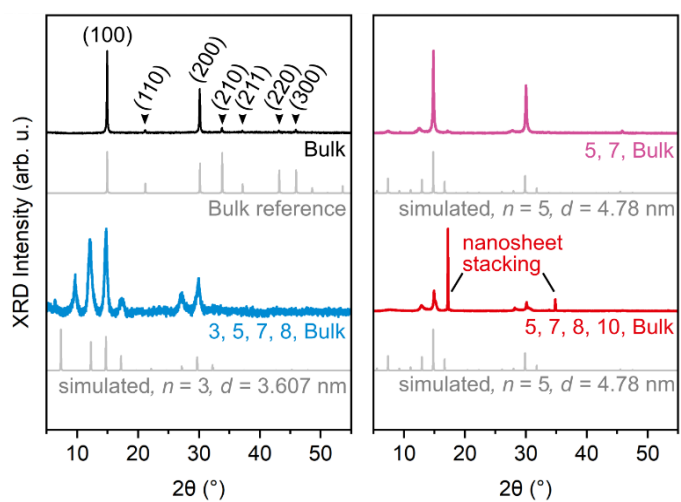

**Figure 2.** Powder XRD patterns of all samples in comparison to calculated diffractograms in grey. **Black:** The bulk sample has a (100) orientation and a respective texture effect (enhancement of (h00) reflexes). (110), (210), (211) etc. are recognized. **Blue:** Layered nanosheets with *n* = 3, 5, 7, 8, and bulk, show a dominant *n* = 3 fraction. **Pink:** *n* = 5, 7, and bulk, with simulation showing a dominant *n* = 5 fraction. **Red:** *n* = 5, 7, 8, 10, and bulk, with simulation showing a dominant *n* = 5 fraction. Reflexes at 17.2° and 34.8° are caused by stacking of the nanosheets, which is confirmed by TEM images (see Fig. S2).

Central to the investigation of transport processes in multi-layered $MAPbBr_3$ nanosheets is whether different layer numbers only occur in separate stacks or if they also coexist within the same perovskite nanosheet, as is schematically shown for $n$ = 3 and 10 in Figure 1f. The scheme also shows an expected defect rich domain at the $n$-$n$ interface inherent to multi-$n$ RPPs, since adjacent layers are shifted by half a unit cell in RPPs.[33] The consequences of this defect-rich domain will be discussed later. The fact that we cannot separate layers by cascade centrifugation gives a first indication that individual nanosheets can contain more than one $n$ . We back up this assumption by spectroscopic data, shown in **Figure 3**. For a sample containing $n$ = 5, 7, 8, 10 and bulk nanosheets (see Fig. 3a), the PL intensity of the bulk PL and the PL for $n$ = 7, when changing the excitation energy is shown (see Figure 3b). For the bulk PL (red curve) in Figure 3b we find excitation features for $n$ = 5, 7, and 8. The presence of these features suggests that excitation of thinner $MAPbBr_3$ regions leads to emission from thicker layers in the same nanosheet, which means that energy or charge transfer must take place by crystal-crystal contacts. When following the PL of $n$ = 7 (violet curve) for a changing excitation energy, excitation features representing the absorbance for $n$ = 5 and 7 get apparent. The crystal-crystal contact of the fractions $n$ = 5 and $n$ = 7 is also evident from this observation. Another argument supporting inter-region contacts is provided by TCSPC PL measurements, shown in Figure S3. PL signals from 2.34 eV to 2.76 eV, show that PL lifetimes get shorter with a decreasing layer number of the nanosheets. The bulk fraction shows longer-lived PL, which indicates that high-$n$ regions in the nanosheets are populated *via* low-$n$ regions by funneling processes, again indicating close crystal-crystal contact.

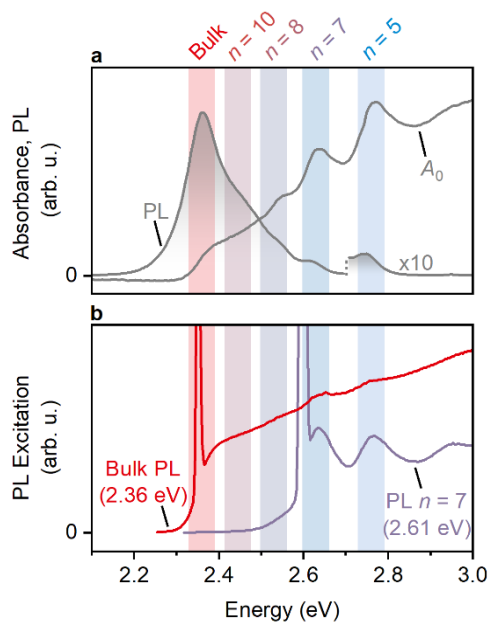

**Figure 3.** Spectroscopic evidence for intercrystal contact in perovskite nanosheets with different layer numbers (for a sample containing $n$ = 5, 7, 8, 10 and bulk nanosheets) in the same perovskite. (**a**) Absorption ($A_0$) and PL spectrum with features caused by different layer

numbers. The main features are marked by colored vertical bars. (**b**) PL signal associated with bulk and *n* = 7 when changing the excitation energy (red and violet, resp.). Both PLE signals show features also present in the absorption spectrum above, which indicates funneling of excited carriers into regions of higher layer numbers.

**2.2. Excitons and Charge Carriers Recombine via Emission and Non-Radiative Auger Processes**

To gain information about the underlying recombination processes in photoexcited MAPbBr$_3$ nanosheets, part of their steady-state absorbance spectra are modelled based on the Elliott theory.[30] The model is derived from studies by Yang *et al.*, Chang *et al.*, and Sestu *et al.*[31,34,35] and described in detail in the SI. The $A_0$ fitting procedure is performed for the energetically lowest transition associated with the highest layer number (bulk). The absorption data and the corresponding model are shown in **Figure 4**a for the bulk transition of MAPbBr$_3$, also containing bulk, 5- and 7-layered regions. The bulk absorption feature contains a charge carrier (green, $A_{CC}$) and an excitonic contribution (orange, $A_{Exc}$). The carrier contribution reflects all transitions of valence band electrons when excited into the conduction band (fundamental absorption), as shown in Figure 4b. The exciton contribution consists of discrete energy states of the bound electron-hole pair, as shown in Figure 4c, representing the exciton as a quasi-particle. In materials with low exciton binding energy such as high-*n* perovskite nanosheets, the excitonic 1s state is populated almost exclusively rather than energetically higher excited states. Up to the absorption shoulder of *n* = 7, the Elliott model accurately represents the absorption profile of multi-layered MAPbBr$_3$ perovskites nanosheets. We extract an exciton binding energy of 20 ± 8 meV and a bandgap of 2.38 ± 0.01 eV attributed to the bulk region. The $E_b$ is less than $k_BT = 26$ meV for 300 K, leading to dissociation of excitons in the bulk. The energetically higher transitions (associated with *n* = 5 and 7, resp.) are clearly visible in the absorption of the sample. These transitions mask $A_{CC}$(bulk) and render its shape for $E > 2.6$ eV speculative. Fitting the energetically higher absorption data to the Elliott model provides an estimate of the exciton binding energy of 136 meV for *n* = 5 (for details see Figure S5). Based on $E_b(n = 5) > k_BT > E_b$(bulk), we can assume that the decay of excitons in low-*n* regions is significantly longer than in bulk regions at room temperature. A strong rise of $E_b$ by thickness confinement is in excellent agreement with literature on MAPbBr$_3$ and other perovskites.[36,37]

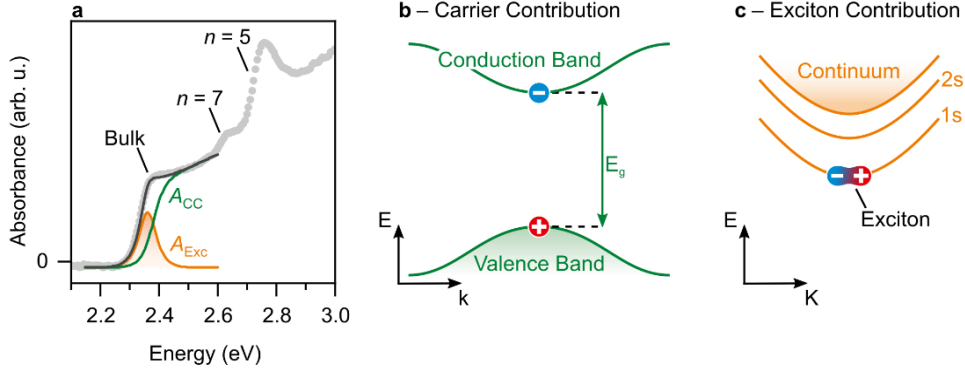

**Figure 4.** Colloidal MAPbBr$_3$ perovskite nanosheet absorption is based on charge carrier and exciton contributions. (**a**) The absorption of MAPbBr$_3$ (*n* = 5, 7, and bulk) shows distinct features, which we assign to the regions of the respective layer numers. Each absorption feature is composed of a step-like charge carrier contribution and a peak-like exciton contribution. (**b**) The charge carrier contribution originates from the band structure. (**c**) The exciton contribution reflects binding states of the excited electron-hole pairs.

**Figure 5**a shows the transient changes in absorption when MAPbBr$_3$ perovskite nanosheets are photoexcited with short laser pulses (< 150 fs) and characterized by TAS. Here, a MAPbBr$_3$ RPP sample containing bulk, 5- and 7-layered fractions was photoexcited with an energy of 2.76 eV and a photon fluence of 50 µJ/cm². We probe a negative ΔA signal at 2.35 eV and a positive signal at 2.45 eV, both decaying over hundreds of picoseconds. The observed transient response is a reported pattern for MAPbBr$_3$ perovskites caused by three factors: (i) band filling, thus fading of the carrier band, (ii) bandgap renormalization (BGR), which causes a shift of the carrier band, and (iii) the presence of free charge carriers, which quench and broaden the excitonic absorption.[34] In Figure 5b, fitting of an extended Elliott model (adapted from Chang *et al.*[31] and Yang *et al.*[34]) is shown (see detailed description in the SI) with excellent agreement to the measured data. In brief, we calculate the excited state absorption with $A^* = A_0 + s_{\Delta A}\Delta A$, with $s_{\Delta A} = 10$ being an empirical scaling factor (further discussed in the SI). The fitting procedure yields time-dependent properties of the electronic structure: the bandgap $E_g$, the exciton binding energy $E_b$, the exciton bandwidth $\Gamma$, the quasi-Fermi energy $E_{fq}$, the carrier temperature $T_e$, and scaling factors $f_1$ and $f_2$, which are proportional to the oscillator strength of the free charge carrier and excitonic transitions, resp. The extended Elliott theory accurately represents the measured ΔA signal and hence serves as a basis for the evaluation of the decay processes in the multi-layered MAPbBr$_3$ perovskite nanosheets studied here.

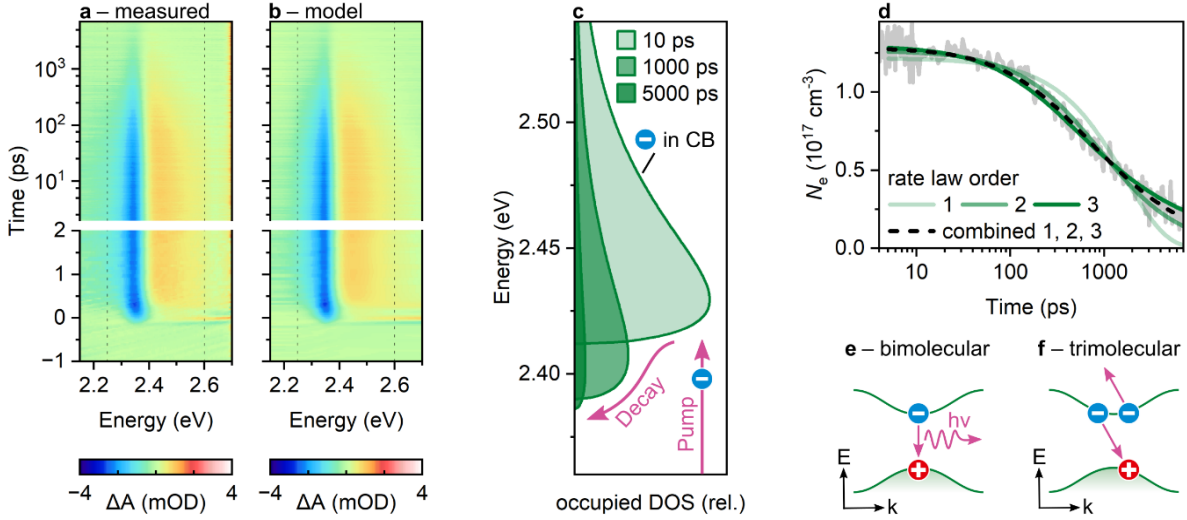

**Figure 5.** Excited charge carriers in MAPbBr$_3$ perovskite nanosheets relax via bi- and trimolecular processes. (**a**) Measured transient ΔA map of MAPbBr$_3$ ($n$ = 5, 7, and bulk) after photoexcitation at 2.76 eV, 50 μJ/cm², and (**b**) corresponding modelled ΔA signal exhibiting excellent agreement (vertical lines are fitting limits). (**c**) Density of populated states in the conduction band (CB) according to the Elliott model show that charge carriers are quasi-Fermi distributed and gradually leave the conduction band. (**d**) The total concentration of excited electrons $N_e$ in the conduction band (grey) is fitted to first, second, and third order kinetics (light to dark green) and to a mixed order kinetic (dashed). $N_e$ is dominated by a second and third order decay indicating bi- and trimolecular processes, schematically depicted in (**e**) and (**f**).

For an intuitive interpretation on how the free charge carriers recombine, the extracted time-dependent properties are translated into a density of occupied states (DOS) for pump-probe delays of 10 ps, 1000 ps, and 5000 ps in Figure 5c. We find that 10 ps after photoexcitation, most of the free charge carriers in the conduction band exhibit an energy of 2.43 eV. After 1000 ps, both the area of the DOS and the bandgap decrease towards the relaxed system. Here, the area corresponds to the number of excited electrons $N_e$ and is calculated from $E_g$ and $E_{fq}$ in the SI. Figure 5d shows that $N_e$ as expected vanishes over time, which is induced by different decay processes: In our experimental set-up the IRF of this measurement is ~150 fs, which means that the excited charge carriers start to thermalize and equilibrate with the lattice temperature during photoexcitation.[38] Which recombination processes occur afterward mainly depends on $N_e$ but also on the sample properties such as the defect concentration in multi-layered MAPbBr$_3$ perovskite nanosheets.[39] In general, higher $N_e$ favors multi-particle recombination. The number of participating charge carriers in turn determines the order of the kinetic rate law by which $N_e$ decays. Figure 5d shows the fits to the rate laws of order 1, 2, and 3 (light to dark green), underpinning that $N_e$ of free charge carriers is only poorly represented by first-order kinetics. This observation is expected, since monomolecular recombination processes in perovskites only dominate at comparatively low charge carrier densities of under $10^{15}$ cm$^{-3}$, which is three orders of magnitude below the fluences applied here.[40] Rate laws of second and third order, on the other hand, better follow the measured data. Second-order kinetics indicate the recombination of free electrons and holes or EEA, while a third order kinetic suggests three-body Auger recombination, as is schematically illustrated in Figure 5e and 4f. The determination of rate constants of the respective orders ($k_1$, $k_2$, and $k_3$) is shown as a dashed line in Figure 5d and calculated using the combined rate law (details given in the SI):

$$\frac{\partial N_e}{\partial t} = -k_1 \cdot N_e - k_2 \cdot N_e^2 - k_3 \cdot N_e^3 \tag{1}$$

Fitting yields rate constants of $k_1 = 1.5 \cdot 10^{-8}$ ps$^{-1}$, $k_2 = 1.2 \cdot 10^{-21}$ cm$^3$/ps, and $k_3 = 1.6 \cdot 10^{-39}$ cm$^6$/ps and respective lifetimes of $\tau_1 = 65$ µs, $\tau_2 = 2.5$ ns, and $\tau_3 = 4.4$ ns for $N_{e,0} = 1.16 \cdot 10^{18}$ cm$^{-3}$. Again, the first order time constant $\tau_1$ being in the microsecond range suggests that monomolecular processes are highly unlikely to occur. In contrast, both $\tau_2$ and $\tau_3$ are in the nanosecond range at this photoexcitation density. Therefore, both Auger and recombination of free charge carriers seem to occur with a rather similar probability. This assumption is supported by the PLQY = 14 % not being quenched by Auger recombination. The relatively slow Auger recombination with respect to spherical nanocrystals was found in CsPbBr$_3$ perovskites as well and is caused by lower collision frequency of charge carriers and a reduced Auger recombination probability per collision.[41] Since the majority of the excited charge carriers are retained for hundreds of picoseconds, we can investigate multi-*n* nanosheets regarding transport processes of both free charge carriers and excitons, which is described in detail in the following section.

## 2.2. Funneling and Exciton Self-Trapping Compete on Picosecond Timescales

To study the role of funneling in multi-*n* perovskite nanosheets, we examine RPPs that respond with multiple signals to photoexcitation. In principle, there are two edge cases for the sequencing in which funneling might take place: (i) Consecutively, by charge carriers and excitons moving in a strictly ascending order through different layer numbers, schematically shown in **Figure 6**a. In this case, longer decays of the transient responses of higher *n* should be observed, when all *n* are excited simultaneously. (ii) Parallel funneling channels exist from all *n* to the most energetically favored state, i.e. to the thickest layer *n* shown in Figure 6b. Transient signals of the thickest nanosheet fraction are expected to exhibit the longest decay times. All other signals should show shorter decays which are similar to one another, if funneling is the fastest decay process. However, we find a mixed form of the described paths occurring, as illustrated in Figure 6c (more detail below). Figure 6d shows the PL (grey) of a sample that contains layer numbers of *n* = 5, 7, 8, 10, and bulk. The sample has the highest PL intensity at 2.45 eV, which means it does not emit preferentially from the layer number with the smallest bandgap. This indicates that funneling from *n* = 10 to bulk is slow in this sample.

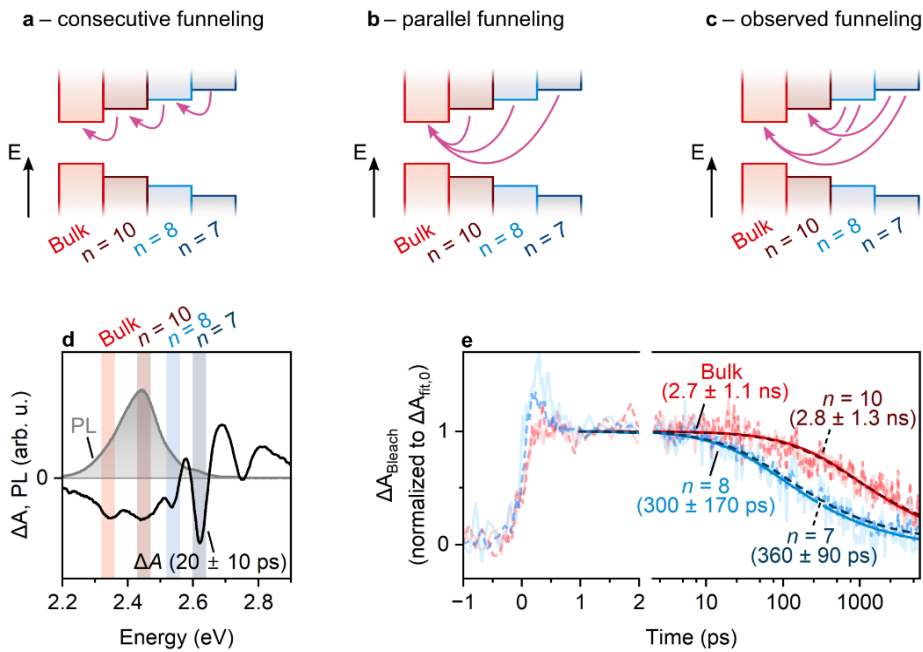

**Figure 6.** Excited charge carriers in MAPbBr$_3$ perovskite nanosheets can funnel via alternative routes into thicker crystal domains. (**a**, **b**, **c**) Schematic band structure of regimes with different layer numbers show how excited electrons funnel on consecutive or parallel pathways, as well as the observed mixed funneling behavior. (**d**) Spectral positions of *n* = 5, 7, 8, 10, and bulk are determined by steady-state PL (grey) and TAS (black, averaged from 10 to 30 ps), with the sample photoexcited at 3.10 eV at a low fluence of 6 μJ/cm². (**e**) Time-resolved amplitudes of the bleach signals in **d** show that carriers from thinner layer number regions (light and dark blue) funnel into regions of higher thickness (light and dark red). Displayed decay times (in brackets) were calculated by fitting to mixed-order decay kinetics.

When photoexciting charge carriers of all *n* simultaneously (at 3.10 eV with 6 μJ/cm²), responses of all layer numbers are visible as negative signals in the ΔA spectrum (black, averaged from 10 to 30 ps). Since we can assign a layer number to each bleach signal, their transients provide information on how excited charge carriers and excitons funnel between different $n$. The normalized $\Delta A$ decays of the different layer numbers are shown in Figure 6e and fitted with rate laws of mixed order (1st, 2nd, and 3rd). The 1/$e$ lifetimes are then determined from the fitted decay curves (for mathematical details see SI). Interestingly, we find two fast and two slow decays. The bleach signals of $n$ = 10 and bulk decay with similar constants of 2.8 ± 1.3 ns and 2.7 ± 1.1 ns, resp., whereas the signals of $n$ = 7 and 8 exhibit shorter similar constants of of 360 ± 90 ps and 300 ± 170 ps, resp. These decays confirm that funneling takes place indeed in an ascending layer number direction. Since $\tau(n = 10) \approx \tau(\text{bulk})$ we conclude that almost no charge carriers funnel from $n = 10$ regimes into the bulk fraction in this sample. This assumption is additionally confirmed by the sample PL centered at 2.45 eV, i.e. the energy associated with $n = 10$. Typically, time constants of charge carrier transfer in perovskites are reported to lie in the picosecond to nanosecond range,[42] which is considerably slower than energy transfer like Förster resonance energy transfer (FRET) taking place in less than a picosecond.[43] With our measured decay times in the picosecond and nanosecond range, we assume that charge carriers are transferred rather than energy.

If mixed-$n$ MAPbBr$_3$ perovskite nanosheets are excited with higher fluences ($\geq$ 50 µJ/cm$^2$), additional interesting photophysics occur, which help to investigate ultrafast charge carrier processes. **Figure 7**a shows a photoinduced absorption (PIA) appearing at 2.2 eV for a colloidal MAPbBr$_3$ nanosheet sample with $n$ = 3, 5, 7, 8, and bulk, which is photoexcited at 3.41 eV. The PIA signal has a rather broad width of 1 eV (see Figure 7b) and, in contrast to the bleach signal at 2.65 eV, cannot be explained *via* the Elliott model. We infer that effects that modify the exciton line shape or change the shape and occupation of the bands are not responsible for the observed PIA. Due to the broad nature of the signal, we assume exciton self-trapping at nonspecific energies, e.g. at edges or defects in the nanosheets. Localized excitons have been suspected in CsPbBr$_3$ and mixed Ag–Sb/Bi perovskites before,[44,45] however to the best of our knowledge, time-resolved studies on trapped excitons in MAPbBr$_3$ perovskites have not been described yet. Lin *et al.* observed a broad PIA for bromide-rich naphthalene diammonium perovskites with $n = 1$ and attributed the feature to a naphthalene triplet species.[46] As opposed to these perovskites, the PIA in the MAPbBr$_3$ nanosheet samples shown here rises over tens of picoseconds, as is shown by the progression of $\Delta A$ at 2.2 eV ± 20 meV in Figure 7c. The PIA follows the kinetic scheme of two consecutive elementary steps (PIA growing and decaying) as depicted by the fit curve (dark blue). The analytical solution to this scheme can be presented as

$$\Delta A_{PIA}(t) = c \cdot \frac{\tau_{grow}}{\tau_{grow} - \tau_{decay}} \cdot \left(e^{-t/\tau_{grow}} - e^{-t/\tau_{decay}}\right), \quad (2)$$

and yields constants of $\tau_{grow} = 16.2 \pm 0.3$ ps, and $\tau_{decay} = 5.97 \pm 0.12$ ns. $\Delta A_{PIA}$ is the time-dependent change of the absorption when excited, and $c$ is a scaling factor. The fact that the $\Delta A$ of the PIA is accurately represented by the described rate law confirms a monomolecular process. Accordingly, excitons and/or charge carriers are trapped with a constant $\tau_{grow}$ and then spontaneously relax with $\tau_{decay}$.

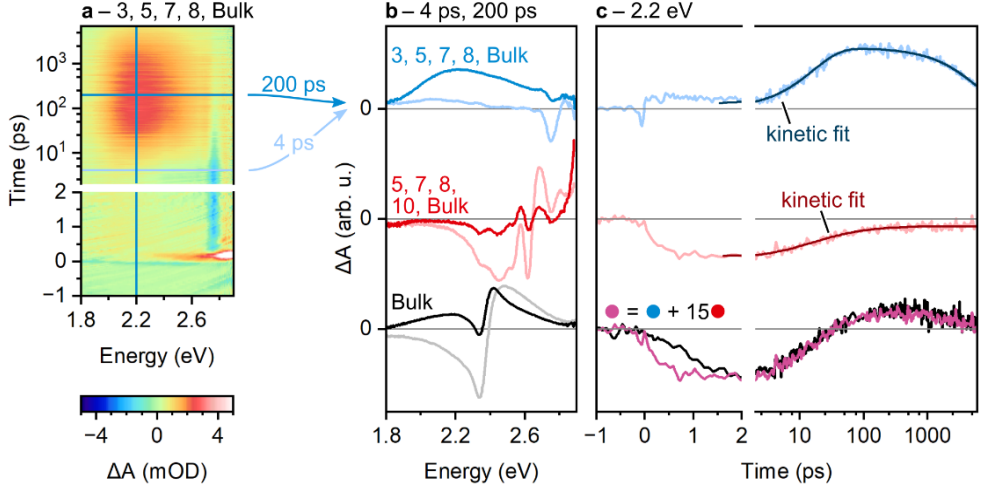

**Figure 7.** Excitons are trapped during funneling. (**a**) $\Delta A$ map of an MAPbBr$_3$ RPP nanosheet sample containing $n$ = 3, 5, 7, 8, and bulk, photoexcited at 3.41 eV with 50 μJ/cm². Besides the bleach signal at 2.75 eV, a broad PIA appears 10 ps after excitation. (**b**) $\Delta A$ spectral linecuts at 4.0 ± 0.4 ps and 200 ± 20 ps (light and dark blue) and linecuts at the same times for a sample containing $n$ = 5, 7, 8, 10, and bulk (red, pump: 3.1 eV, 50 μJ/cm²), and a sample of bulk only (black, pump: 3.41 eV, 50 μJ/cm²). (**c**) Corresponding $\Delta A$ traces at 2.2 eV ± 20 meV with kinetic fits and the mirroring of the response of the bulk perovskite by combining the other two measurements shows that exciton trapping and gain are competing processes in multi-$n$ MAPbBr$_3$ RPP nanosheets.

In comparison, for a sample with $n$ = 5, 7, 8, 10, and bulk, photoexcitation does not lead to the broad PIA described above, as is illustrated by the spectral cuts and transients in red in Figure 7b and c. The sample is photoexcited with a fluence of 50 μJ/cm² at an energy of 3.1 eV, generating excitons and free charge carriers in each layer number, which is supported by the respective narrow bleach signals. The absence of a broad PIA signal implies that excitons are not trapped in this sample. Instead, a negative $\Delta A$ signal is observed, which can be caused by gain, decays with $\tau = 27.7 \pm 2.6$ ps and is broadened with respect to the measured PL. A high PLQY of 14 % compared to the RPP sample with 3.6 % ($n$ = 3, 5, 7, 8, and bulk) also indicates that excitons are substantially less trapped and then recombine non-radiatively in samples with a better *n-n* connectivity. The crystalline quality and surface passivation of multi-layered MAPbBr$_3$ perovskite nanosheets plays a crucial role for efficient PL since a high number of free electrons is needed for optical gain, as has been shown for MAPbI$_3$ perovskites.[47] Consequently, both, the broad PIA and gain, can be observed in parallel when stimulated emission occurs before charge carriers are trapped in localized states. In Figure 7b and c in black, this observation is made on MAPbBr$_3$ perovskites, which contain bulk-like layer numbers only. The temporal evolution of ΔA can be recreated as a linear combination of the two cases, where either a broad PIA or gain is present (pink in Figure 7c). This gives solid evidence that multiple factors, such as the combination of *n*, defect density, and the *n-n* connectivity have influence on whether exciton-trapping or gain is dominating in the colloidal samples.

## 3. Conclusion

In this work we have addressed recombination (pathways) and funneling processes in colloidal multi-layered 2D MAPbBr$_3$ perovskite nanosheets. By applying an extended Elliott model we find that after photoexcitation, second and third order processes dominate recombination in the nanosheets. Funneling in MAPbBr$_3$ nanosheets can occur, when thinner layer numbers are connected with higher-*n* layers. Long lived low-*n* excitons ($E_\mathrm{b} >$ 100 meV) funnel into high-*n* layers with $\tau =$ 10 ps to 50 ps, where their $E_\mathrm{b}$ is drastically decreased below $k_\mathrm{B}T = 26$ meV at room temperature. Depending on the connectivity of the layers with different *n*, the excitons and charge carriers funnel into the direction of increasing thickness, but can accumulate in intermediate *n*, where they recombine radiatively or by Auger processes. Additionally, the quality of the *n-n* interface determines whether exciton trapping occurs, and thus competes with funneling. This study shows that funneling and self-trapping processes in RPPs are key parameters for designing energetic pathways for excitons and free charge carriers for future efficient RPP-based devices. Structure, quality and size of *n-n* interfaces define the kinetics of trapping processes in mixed-n perovskite nanosheets and is subject of further synthetic improvement. Funneling in the nanosheets exhibits mixed character (consecutive and parallel) and allows for different charge transfer and transport configurations. When comparing mixed-*n* nanosheets with their bulk-like counterparts, we envision application possibilities for the nanosheets toward directional funneling and exciton binding energy downscaling, which could help to outcompete trapping and EEA.

## 4. Methods

*Steady-State Spectroscopy:* UV/vis absorption spectra were obtained with a Lambda 1050+ spectrophotometer from Perkin Elmer equipped with an integration-sphere. The PL measurements were performed with a fluorescence spectrometer (Spectrofluorometer FS5, Edinburgh Instruments). PLQYs of the samples were determined using an absolute method by directly exciting the sample solution and the reference (toluene in our case) in an SC-30 integrating sphere module fitted to the Spectrofluorometer FS5 from Edinburgh Instruments. During the measurement, the excitation slit was set to 6.5 nm, and the emission slit was adjusted to obtain a signal level of $10^6$ cps, the excitation wavelength was chosen at 375 nm with 50 μW/cm² irradiance. A wavelength step size of 0.1 nm and an integration time of 0.2 s was used. The calculation of absolute PLQY is based on the formula, $\eta = \frac{E_{\text{sample}} - E_{\text{ref}}}{S_{\text{ref}} - S_{\text{sample}}}$, where $\eta$ is the absolute PLQY, $E_{\text{sample}}$ and $E_{\text{ref}}$ are the integrals at the emission region for the sample and the reference, resp., and $S_{\text{sample}}$ and $S_{\text{ref}}$ are the integrals at the excitation scatter region for the sample and the reference, respectively. The selection and calculation of integrals from the emission and excitation scattering region and the calculation of absolute PLQY were performed using the FLUORACLE software from the Edinburgh Instruments.

*Transient Absorption Spectroscopy (TAS):* Ultrafast carrier dynamics of colloidal MAPbBr$_3$ perovskite nanosheets were studied by femtosecond transient absorption spectroscopy. The samples were dispersed in toluene under inert gas conditions and measured inside a 2 mm quartz glass cuvette with constant stirring. Laser pulses (120 fs, 1.54 eV or 90 fs, 1.55 eV) with a repetition rate of 1 kHz are generated by a Ti:sapphire regenerative amplifier systems (Astrella-F from Coherent or Spitfire ACE from Spectra-Physics) and split 90:10 into pump and probe pulses. The pump beam is led into an optical parametric amplifier and second harmonics generation system (Appollo-T by Ultrafast Systems or TOPAS Prime by Spectra-Physics), where its energy is adjusted by nonlinear frequency mixing to the desired excitation energy between 0.77 eV and 4.28 eV. The probe pulse is transformed into a broadband supercontinuum by nonlinear processes in a CaF$_2$ or sapphire crystal. Accordingly, probing takes place in an energy range from 1.55 eV to 3.87 eV. The overlap of pump and probe pulses within the sample is optimized before each measurement to achieve the best signal-to-noise ratio. The attenuated probe pulse is led into a fiber-coupled spectrometer (Ultrafast, Helios Fire) with a linear detector array. The charge carrier dynamics via the absorption changes are studied by varying the delay times between probe and pump pulse. The pump-probe delay is electronically controlled with a delay line and covers a time window of up to 8 ns after photoexcitation.

*Transmission Electron Microscopy (TEM):* TEM samples were prepared by diluting the nanosheet suspension with toluene followed by drop casting 10 μL of the suspension on a TEM copper grid coated with a carbon film. Standard images were done on a Talos-L120C and EM-912 Omega with a thermal emitter operated at an acceleration voltage of 120 kV and 100 kV.

*X-ray diffraction (XRD):* Measurements were performed on a Panalytical Aeris System with a Bragg-Brentano geometry and a copper anode with an X-ray wavelength of 0.154 nm from the Cu-kα1 line. The samples were measured by drop-casting the suspended samples on a ⟨911⟩ or ⟨711⟩ cut silicon substrate. Powder XRD pattern are simulated with the PowderCell package[48], Version 2.4. The inorganic part of the crystal (Pb-Br) unit is used to simulate the pattern with Rietveld-Toraya (plate) preferred orientation in [001] direction based on the crystallographic information from Elbaz *et al.*[32] Unit cells are formed with different *n* from 2 to 6 in Ruddlesden-Popper stacking configuration.

**Supporting Information**
Supporting Information is available from the Wiley Online Library or from the author.

**Acknowledgements**
DFG is acknowledged by C. K. for funding of SFB 1477 "Light-Matter Interactions at Interfaces", project number 441234705, W03 and W05. C. K. also acknowledges the European Regional Development Fund of the European Union for funding the PL spectrometer (GHS-20-0035/P000376218) and X-ray diffractometer (GHS-20-0036/P000379642) and the DFG for funding an electron microscope Jeol NeoARM TEM (INST 264/161-1 FUGG) and an electron microscope ThermoFisher Talos L120C (INST 264/188-1 FUGG). J. L. gratefully acknowledges funding by the Deutsche Forschungsgemeinschaft (DFG, German Research Foundation) under the Excellence Strategy of the Cluster of Excellence PhoenixD (EXC 2122, Project ID 390833453) and by the Ministry for Science and Culture of the State of Lower Saxony (MWK) for a Stay Inspired: European Excellence for Lower Saxony (Stay-3/22-7633/2022) Grant. J. L. is also grateful for funding by an Athene Grant of the University of Tübingen (by the Federal Ministry of Education and Research (BMBF) and the Baden-Württemberg Ministry of Science as part of the Excellence Strategy of the German Federal and State Governments). TA measurements were funded by the DFG under contract INST 37/1160-1 FUGG (project nr. 458406921). The authors acknowledge access to the Ti:sapphire amplifier system (major equipment DFG, Project ID 231415720).

# Supporting Information
# Understanding the Optoelectronic Processes in Colloidal 2D Multi-Layered MAPbBr$_3$ Perovskite Nanosheets: Funneling, Recombination and Self-Trapped Excitons

André Niebur[1,2], Eugen Klein[3], Rostyslav Lesyuk[3,4], Christian Klinke[3,5,6], and Jannika Lauth*[1,2,7,8]

[1]*Institute of Physical Chemistry and Electrochemistry, Leibniz University Hannover, Callinstr. 3a, D-30167 Hannover, Germany*
[2]*Cluster of Excellence PhoenixD (Photonics, Optics, and Engineering – Innovation Across Disciplines), Hannover, Germany*
[3]*Institute of Physics, University of Rostock, Albert-Einstein-Straße 23, 18059 Rostock, Germany*
[4]*Pidstryhach Institute for Applied Problems of Mechanics and Mathematics of NAS of Ukraine, Naukowa Str. 3b, 79060 Lviv*
[5]*Department Life, Light & Matter, University of Rostock, Albert-Einstein-Strasse 25, 18059 Rostock, Germany*
[6]*Department of Chemistry, Swansea University – Singleton Park, Swansea SA2 8PP, United Kingdom*
[7]*Laboratory of Nano and Quantum Engineering (LNQE), Leibniz University Hannover, Schneiderberg 39, D-30167 Hannover, Germany*
[8]*Institute of Physical and Theoretical Chemistry, University of Tübingen, Auf der Morgenstelle 18, D-72076 Tübingen, Germany*

*E-mail: jannika.lauth@uni-tuebingen.de

## 1. Methods

### 1.1. Chemicals and reagents

All chemicals were used as received: Lead(II) acetate tri-hydrate (Aldrich, 99.999%), nonanoic acid (Alfa Aesar, 97%), tri-octylphosphine (TOP; ABCR, 97%), methylammonium bromide (MAB; Aldrich, 98%), diphenyl ether (DPE; Aldrich, 99%), toluene (VWR, 99,5%), dimethylformamide (DMF; Aldrich, 99,8%), 1-bromotetradecane (BTD; Aldrich, 97%), dodecylamine (DDA; Merck, 98%).

### 1.2. PbBr$_2$ nanosheet synthesis[1]

In a typical synthesis a three neck 50 mL flask was used with a condenser, septum and thermocouple. 860 mg of lead acetate tri-hydrate (2.3 mmol) were dissolved in 10 mL of nonanoic acid (57 mmol) and 10 mL of 1-bromotetradecane (34 mmol) and heated to 75 °C until the solution turned clear in a nitrogen atmosphere. Then vacuum was applied to remove the acetic acid which is generated by the reaction of nonanoic acid with the acetate from the lead precursor. After 1.5 h the reaction apparatus was filled with nitrogen again and the reaction was started by adding 0.06 mL of TOP (0.13 mmol) at a temperature of 140 °C and was stopped 8 min later. The solution was left to cool down below 80 °C. Afterwards, it was centrifuged one time at 4000 rpm for 3 min. The particles were suspended in 7.5 mL of toluene and put into a freezer for storage.

### 1.3. Synthesis of bulk-like MAPbBr$_3$ nanosheets

A three neck 50 mL flask was used with a condenser, septum and thermocouple. 10.5 mL of diphenyl ether (66.6 mmol), 0.15 mL of a 500 mg dodecylamine (2.70 mmol) in 4 mL diphenyl ether precursor and 0.2 mL (0.44 mmol) of TOP were heated to 80 °C in a nitrogen atmosphere. Then vacuum was applied to dry the solution. After 1 h the reaction apparatus was filled with nitrogen again, the temperature was increased to 120 °C and 1.5 mL of as prepared PbBr$_2$ nanosheets in toluene were added. The synthesis was started with the injection of 0.1 mL of a 300 mg methylammonium bromide (2.68 mmol) in 6 mL dimethylformamide precursor. After 15 min the heat source was removed and the solution was left to cool down below 60 °C. Afterwards, it was centrifuged at 4000 rpm for 3 min. The particles were washed two times in toluene before the product was finally suspended in toluene again and put into a freezer for storage.

### 1.4. Synthesis of MAPbBr$_3$ nanosheets with *n* = 3, 5, 7, 8, bulk

A three neck 50 mL flask was used with a condenser, septum and thermocouple. 10.5 mL of diphenyl ether (66.6 mmol), 0.2 mL of a 500 mg dodecylamine (2.70 mmol) in 4 mL diphenyl ether precursor and 0.2 mL (0.44 mmol) of TOP were heated to 80 °C in a nitrogen atmosphere. Then vacuum was applied to dry the solution. After 1 h the reaction apparatus was filled with nitrogen again, the temperature was increased to 160 °C and 2.5 mL of as prepared PbBr$_2$ nanosheets in toluene were added. The synthesis was started with the injection of 0.03 mL of a 300 mg methylammonium bromide (2.68 mmol) in 6 mL dimethylformamide precursor after all of the PbBr$_2$ nanosheets were dissolved. The reaction mixture was cooled down to 60 °C in a timeframe of 35 min after all of the initially formed MAPbBr$_3$ at 160 dissolved. The desired nanosheets started to form at 90 °C. Afterwards, it was centrifuged at 4000 rpm for 3 min. The particles were washed two times in toluene before the product was finally suspended in toluene again and put into a freezer for storage.

### 1.5. Synthesis of MAPbBr$_3$ nanosheets with *n* = 5, 7, bulk

A three neck 50 mL flask was used with a condenser, septum and thermocouple. 10.5 mL of diphenyl ether (66.6 mmol), 0.3 mL of a 500 mg dodecylamine (2.70 mmol) in 4 mL diphenyl ether precursor and 0.2 mL (0.44 mmol) of TOP were heated to 80 °C in a nitrogen atmosphere. Then vacuum was applied to dry the solution. After 1 h the reaction apparatus was filled with nitrogen again, the temperature was increased to 120 °C and 1.5 mL of as prepared PbBr$_2$ nanosheets in toluene were added. The synthesis was started with the injection of 0.1 mL of a 300 mg methylammonium bromide (2.68 mmol) in 6 mL dimethylformamide precursor. After 10 min the heat source was removed and the solution was left to cool down below 60 °C. Afterwards, it was centrifuged at 4000 rpm for 3 min. The particles were washed two times in toluene before the product was finally suspended in toluene again and put into a freezer for storage.

### 1.6. Synthesis of MAPbBr$_3$ nanosheets with *n* = 5, 7, 8, 10, bulk

A three neck 50 mL flask was used with a condenser, septum and thermocouple. 10.5 mL of diphenyl ether (66.6 mmol), 0.2 mL of a 500 mg dodecylamine (2.70 mmol) in 4 mL diphenyl ether precursor and 0.2 mL (0.44 mmol) of TOP were heated to 80 °C in a nitrogen atmosphere. Then vacuum was applied to dry the solution. After 1 h the reaction apparatus was filled with nitrogen again, the temperature was increased to 120 °C and 0.8 mL of as prepared PbBr$_2$ nanosheets in toluene were added. The reaction temperature was reduced to 80 °C after all of the PbBr$_2$ dissolved. The synthesis was started with the injection of 0.06 mL of a 300 mg methylammonium bromide (2.68 mmol) in 6 mL dimethylformamide precursor. After the injection the temperature was increased to 120 °C. After 10 min the heat source was removed and the solution was left to cool down below 60 °C. Afterwards, it was centrifuged at 4000 rpm for 3 min. The particles were washed two times in toluene before the product was finally suspended in toluene again and put into a freezer for storage.

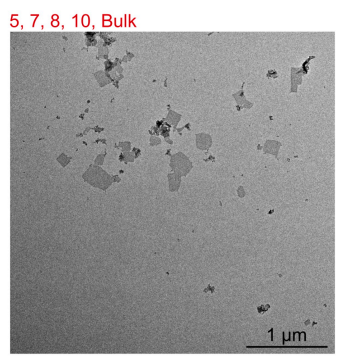

**Figure S1.** TEM image of a sample with $n = 5, 7, 8, 10$, and bulk fractions, with a cutout of the same size as other TEM images in Figure 1.

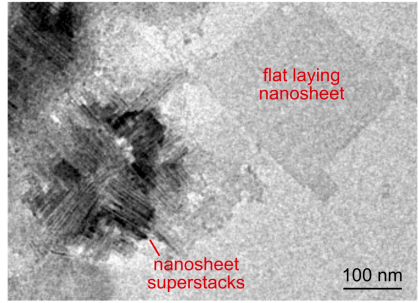

**Figure S2.** TEM image of a sample with $n = 5, 7, 8, 10$, and bulk fractions. A flat laying nanosheet (right) and nanosheet superstacks, seen from the side (left).

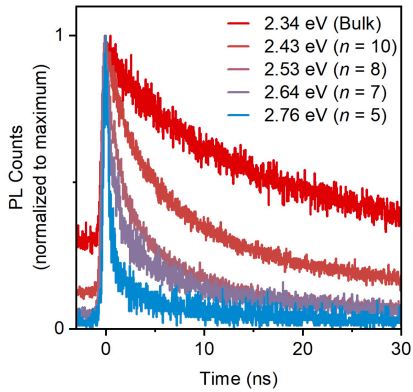

**Figure S3.** Time-resolved (TCSPC) PL decay from 2.34 eV to 2.76 eV (red to blue), showing that PL associated with lower layer numbers is decaying faster than PL of higher layer numbers. The low-$n$ regions demonstrate substantially faster decays (starting from subnanoseconds) than higher-$n$ regions. The bulk domain has the longest PL lifetime with a time constant of nearly 16 ns. This evidence supports the picture of cascade population of higher-$n$ regions on cost of the depopulation of lower ones.

## 2. Elliott Formula and least square fitting

The model for absorption data as proposed by Elliott in 1957[2] was applied following Yang et al.[3] and Chang et al.[4] who have extended the model to excited state absorption. We have adapted the model as follows.

### 2.1. Ground State Absorption

The Elliott theory puts the properties of the band structure and the excitons in a semiconductor in relation to its absorption. In principle, the absorption is obtained by adding two contributions: A charge carrier contribution $A_{CC}$ and an exciton contribution $A_{Exc}$:

$$A_0 = A_{CC} + A_{Exc} \tag{S1}$$

The charge carrier contribution is calculated by integrating over all possible transitions from the valence band to the conduction band as

$$A_{CC} = f_1 \frac{2\pi\sqrt{E_b}}{\hbar\omega} \frac{1}{\Gamma} \int_{E_g}^{\infty} \text{sech}\left(\frac{\hbar\omega - E}{\Gamma}\right) \frac{\zeta(E - E_g)}{1 - \exp\left(-2\pi\sqrt{\frac{E_b}{E - E_g}}\right)} dE \tag{S2}$$

with $f_1$ a scaling factor proportional to the oscillator strength, $E_b$ the exciton binding energy, $\hbar\omega$ the photon energy, $\Gamma$ the exciton bandwidth, $E_g$ the band gap and sech the hyperbolic secant function. $\zeta$ accounts for the non-parabolic shape of the valence and conduction band,

$$\xi(E - E_g) = 1 + 10R \cdot E + 126R^2 \cdot E^2 \tag{S3}$$

with $R$ the non-parabolic factor ($R = m^2/\hbar^4 c_{np}$) for a dispersion relation of

$$E(k) = \frac{\hbar^2}{2m}k^2 - c_{np}k^4 \tag{S4}$$

Setsu *et al.* reported $R = 0.091$ eV$^{-1}$ for MAPbBr$_3$.[5] In our case we found optimal fits for $R = 0.049$ eV$^{-1}$ for the bulk sample. The exciton contribution $A_{Exc}$ is the sum of the excitonic states with binding energies of the $j$-th excitonic state of $E_{b,j} = E_b/j^2$:

$$A_{Exc} = f_2 \frac{2\pi (E_b)^{3/2}}{\hbar\omega} \frac{1}{\Gamma} \sum_{j=1}^{7} \frac{1}{j^3} \text{sech}\left(\frac{\hbar\omega - E_g + E_b/j^2}{\Gamma}\right) \tag{S5}$$

with $f_2$ a scaling factor proportional to the oscillator strength. We consider the first seven excitonic states ($1 \leq j \leq 7$), as only the first or the first two states have a significant contribution to the overall absorption. For the MAPbBr$_3$ perovskite only containing bulk-like regions, the optimized fitting curve is shown in Figure S4, with the optimized parameters in Table S1.

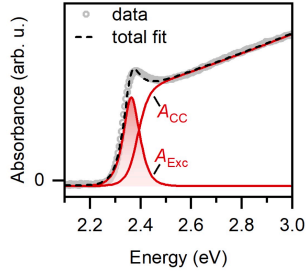

**Figure S4.** Bulk-like MAPbBr$_3$ perovskite is modelled according to Elliott theory. The exciton contributions (filled) and charge-carrier contributions (not filled) are displayed, respectively.

**Table S1.** Fitting limits ($E_{min} \ldots E_{max}$), optimized parameters and $R^2$ of bulk-like MAPbBr$_3$ in Figure S4. $A_0$ is fitted to Eq. (S1).

| | |
|---:|:---:|
| $E_{min} \ldots E_{max}$ (eV) | 2.0 ... 3.0 |
| $E_g$ (eV) | 2.389 ± 0.003 |
| $E_b$ (meV) | 28.2 ± 2.7 |
| $\Gamma$ (meV) | 27.5 ± 0.4 |
| $f_1$ | 0.66 ± 0.03 |
| $f_2$ | 1.55 ± 0.17 |
| $R$ (eV$^{-1}$) | 0.0490 ± 0.0026 |
| offset | -0.0437 ± 0.0022 |
| $R^2$ | 0.9992 |

For a MAPbBr$_3$ perovskite containing $n = 5, 7$, and bulk, the absorbance has to be fitted as $A_0 = A_0(\text{bulk}) + A_0(n = 7) + A_0(n = 5)$. As the absorption features of $n = 5$ and $n = 7$ are close to eachother, we first optimize $A_0(\text{bulk})$ and then optimize $A_0(n = 7)$ and $A_0(n = 5)$ simultaneously. The fitted curves are shown in Figure S5 and the corresponding parameters are in Table S2.

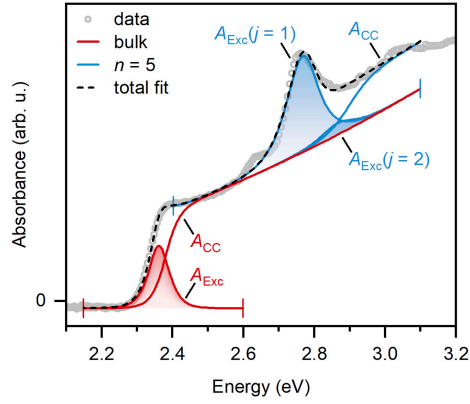

**Figure S5.** Multi-thickness ($n = 5, 7$, and bulk) MAPbBr$_3$ perovskite is modelled according to Elliott theory. The exciton contributions (filled) and charge-carrier contributions (not filled) are displayed in blue and red for $n = 5$ and bulk, respectively. The fitting parameters are in Table S2.

**Table S2.** Fitting limits ($E_{min} \ldots E_{max}$), optimized parameters and $R^2$ of the respective fits. First, the feature of $A_0$ corresponding to the bulk fraction is fitted to Eq. (S1). $A_0$ is then substracted, by the fitted bulk absorption. This difference is fitted to the Elliott equation representing the spectral feature of the fraction $n = 5$. The $n = 7$ fraction is neglected, due to its small influence on the absorption.

|  | Bulk | $n = 5$ |
|---:|:---:|:---:|
| $E_{min} \ldots E_{max}$ (eV) | 2.15 … 2.55 | 2.40 … 3.1 |
| $E_g$ (eV) | 2.37 ± 0.01 | 2.907 ± 0.004 |
| $E_b$ (meV) | 16 ± 8 | 136 ± 4 |
| $\Gamma$ (meV) | 23.5 ± 0.6 | 44.8 ± 1.5 |
| $f_1$ | 0.39 ± 0.09 | 0.144 ± 0.005 |
| $f_2$ | 0.8 ± 0.4 | 0.161 ± 0.006 |
| $R$ (eV$^{-1}$) | 0.024 ± 0.038 | 0.030 ± 0.004 |
| offset | -0.0278 ± 0.0009 | 0 (fixed) |
| $R^2$ | 0.9995 | 0.991 |

## 2.2. Excited State Absorption

For a semiconductor in its excited state, the absorption is given as

$$A^* = A^*_{CC} + A^*_{Exc} \tag{S6}$$

with $A^*_{CC}$ and $A^*_{Exc}$ being the charge carrier contribution and the exciton contribution of the pumped system. The excited charge carrier contribution to the absorption is

$$A^*_{CC} = f_1 A_1 \frac{2\pi \sqrt{E^*_b}}{\hbar \omega} \frac{1}{\Gamma^*} \int_{E^*_g}^{\infty} \mathrm{sech}\left(\frac{\hbar\omega - E}{\Gamma^*}\right) \frac{\xi(E - E^*_g)}{1 - \exp\left(-2\pi \sqrt{\frac{E^*_b}{E - E^*_g}}\right)} dE \cdot (1 - f_{FD})^2 \tag{S7}$$

which is similar to formula (S2) for the absorption of the ground state $A_{CC}$. Photoinduced changed properties are $E^*_g$, $E^*_b$, and $\Gamma^*$, derived from their ground state counterparts $E_g$, $E_b$, and $\Gamma$, respectively. $f_1$ is a scaling factor for the transition dipole moment $f_1$ after a photoinduced change. The occupation of transitions after excitation is considered by the Fermi distribution $f_{FD}$, leading to Pauli exclusion of these transitions, and is calculated by

$$f_{FD}(\omega) = \frac{1}{1 + \exp\left(\frac{\hbar\omega - E_{fq}}{k_B T_e}\right)} \tag{S8}$$

with $E_{fq}$ the quasi-Fermi energy and $T_e$ the temperature of the excited charge carriers (both electrons and holes), which we set to 400 °C. The exciton contribution of the pumped system is

$$A^*_{Exc} = f_2 A_2 \frac{2\pi \left(E_b^*\right)^{3/2}}{\hbar\omega} \frac{1}{\Gamma^*} \sum_{j=1}^{7} \frac{1}{j^3} \text{sech}\left(\frac{\hbar\omega - E_g^* + E_b^*/j^2}{\Gamma^*}\right) \tag{S9}$$

with $f_2$ a scaling factor for the transition dipole moment $A_2$. In this work, the ground state absorption is fitted by a least square approach to the Elliott theory as in formula (S1), yielding the ground state parameters $E_g$, $E_b$, $\Gamma$, $A_1$, $A_2$, $R$ and a constant offset as a background. The excited state absorption is then calculated by

$$A^* = A_{0,\text{fitted}} + s_{\Delta A} \cdot \Delta A_{\text{measured}} \tag{S10}$$

with $s_{\Delta A}$ an empirical scaling factor to put the ground state absorption $A_{0,\text{fitted}}$ and the measured transient absorption data $\Delta A_{\text{measured}}$ into relation, as both quantities derive from different devices. By scaling $\Delta A_{\text{measured}}$ we also take into account, that not each probed MAPbBr$_3$ nanoparticle was excited by the pump laser. In Figure S6 the measured ground state absorption ($A_0$, dashed), the measured absorption change ($\Delta A$, red) for $t = (1.0 \pm 0.1)$ ps and excited state absorption $A^*$ for $s_{\Delta A}$ from 10 to 500 are shown (light green to blue).

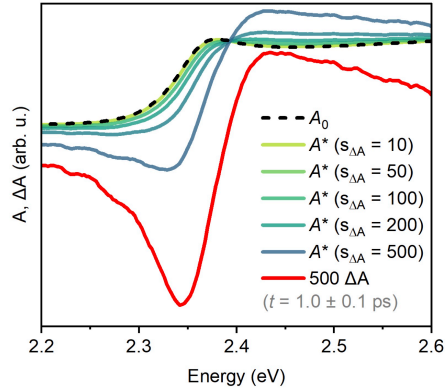

**Figure S6.** Estimation of $s_{\Delta A}$. Measured ground state absorption (dashed) and $\Delta A$ (red) of a sample containing $n = 5, 7$, and bulk. The excited state absorption is the sum of $A_0$ and $s_{\Delta A} \cdot \Delta A$ displayed for various $s_{\Delta A}$ (yellow to blue).

We then fit $A^*$ for different empiric scaling factors $s_{\Delta A}$ to follow how the fitting parameters change. The coefficient of determination $R^2$ and the fitting parameters $E_b$, $E_g$, $\Gamma$, $E_{fq}$, $f_1$, $f_2$ and the offset are plotted against $s_{\Delta A}$ in Figure S7, respectively.

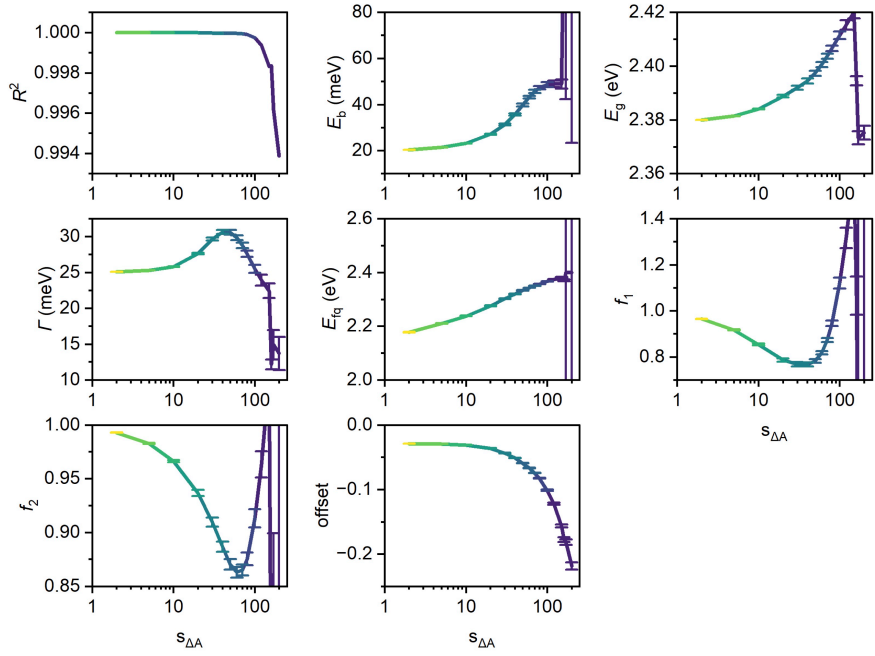

**Figure S7.** Estimation of $s_{\Delta A}$. Fitting parameters and $R^2$ for the extended Elliott model of a sample containing $n = 5, 7$, and bulk for $t = 1 \pm 0.1$ ps.

We find, that $R^2$ significantly decreases for $s_{\Delta A} > 100$, which is a sign for the model poorly describing the data in that range. Additionally, the errors of the fitting parameters drastically increase for $s_{\Delta A}$ greater than 100 and randomize at 200. Therefore, we aim for smaller $s_{\Delta A}$.

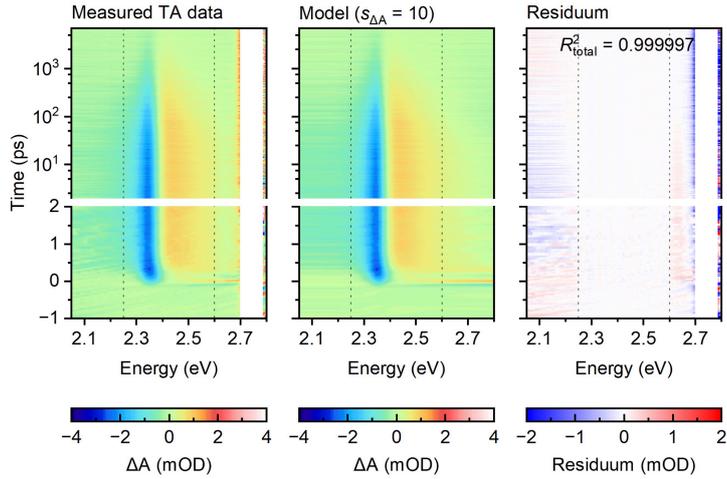

**Figure S8.** Measured $\Delta A$ map of a sample containing fractions with $n = 5$, $n = 7$, and bulk, pumped at 2.76 eV (left), the modelled data following the extended Elliott model (middle), and the residuum (right). The measured data is well represented by the model in the fitting limits of 2.25 eV to 2.6 eV, confirming the validity of using the Elliott model here.

## 2.3. Number of Excited Electrons and Holes

We calculate the number of electrons following the procedure by Chang *et al.*[4] In short, equation S11 is used to determine $N_\text{e}$, which corresponds to the area of the occupied density of states of the conduction band. The number of excited electrons and excited holes are considered to be equal.

$$N_\text{e} = \int_{E_\text{g}}^{\infty} \frac{1}{2\pi^2} \left(\frac{2m_\text{eff}}{\hbar^2}\right)^{3/2} \sqrt{E - E_\text{g}^*} \, f_\text{FD} \, dE \tag{S11}$$

with $E_\text{g}^*$ and $f_\text{FD}$ originating from the fitting procedure of $A^*$. For simplicity, the effective mass of the electron and hole $m_\text{eff}$ is assumed to be $m_\text{e}$, the invariant mass of the electron.